\documentclass[article%
 reprint,%
 amssymb, amsmath,%
 aip,cha,%
]{revtex4-1}
\usepackage{graphicx}
\usepackage{framed}

\usepackage{setspace}
\usepackage{fancyvrb}

\usepackage[hyphenbreaks]{breakurl}
\usepackage{hyperref}
\fvset{formatcom=\singlespacing}

\begin{document}

\section*{A no-phone/no-app contact tracing hardware token}
\noindent {\sl T. Bensky, California Polytechnic State University, Department of Physics, San Luis Obispo, CA, 93407, tbensky@calpoly.edu}

\section*{Abstract}

We report the development of an open-source, hardware-based contact tracer, made from readily available parts, costing less than $\$20$ USD.  This work was motivated by the need for a technology-assisted contact tracer that avoids privacy issues found with those involving a mobile phone. Contact tracing is done here {\sl without the use of a mobile phone or an app at all}. Instead, contact tracing is implemented using Bluetooth Low Energy on an ESP32 micro-controller.  The ESP32 is used to both advertise and receive health information to others in close proximity, forming a strictly peer-to-peer contact tracer.  The contact tracer can be assembled by an individual and configured use within minutes.

\section{Introduction}

Contact tracing (CT) has long been known to help slow the spread of infectious diseases.~\cite{ct}   In our modern era, mobile phones could be an ideal technology to streamline and automate CT, but this has not proven to be the case.  The app landscape in this regard is greatly fragmented,~\cite{frag} and has shown an ongoing litany of privacy and security issues.~\cite{app_security}  It is also becoming clear that CT, either with or without available apps is not being maximally effective in slowing the current pandemic.~\cite{ct_woes} In this work, we discuss a different approach.

Here we demonstrate a hardware device, capable of contact tracing, that does not use a mobile phone or an app at all. Instead, it uses a low-cost ESP32 micro-controller,~\cite{esp32} popular with the ``maker'' community.  In use, it is to be maintained and carried by an individual as they go about their needs in public, presumably around other people.  There is precedence for such a ``hardware tracing token.''~\cite{hardware_token}

The ESP32 (hereafter the ``token''), used as a contact tracer here is programmed to exchange (send and receive) health information with those in close proximity (i.e.``encounters'') who are also carrying the token.  Bluetooth Low Energy (BLE) is used for this exchange. Setting health information is strictly between the user and the token via configuration software, which is open-source, and may be easily scrutinized for privacy and security claims.  The software is used locally on the user's computer, free of any network, server or centralized database access.   The same goes for receiving health information from encounters.  

All data in this CT system is wholly contained within the user/token ecosystem and is secured by hardware restriction, also known as hardware DRM.~\cite{drm} In this case, hardware restrictions are that WiFi on the token is disabled, the device stores encounters only as long as it is powered, has no outside access points (i.e. network, keyboard, etc.), only stores an anonymized identification to a given user, and deals with health issues that are only a list of user-selected symptoms, which are not tied to any formal health records.

In the following sections, the construction and operation of the token is presented, followed by a discussion of ``self-motivated'' CT this system requires, then some conclusions.  All details and resources for this project can be found at \url{https://github.com/tbensky/npct}.

\section{Construction and preparation for use}

\subsection{Construction}

Constructing this device has a ``maker'' theme and can be done within minutes by an individual wishing to participate in CT.  It consists of three parts, all readily available (some of which the individual may already have).  The first is the token (the ESP32 micro-controller), which has an average price of around \$8 USD and can be purchased from a variety of sources.~\cite{esp32_buy}   The second is a battery to power the token.  This is most conveniently a ``USB battery'' typically used to provide extra mobile-phone power (i.e. a battery power-pack that can be used to charge a mobile phone).~\cite{battery} Lastly, one needs a cable to connect the battery to the token. The token has a micro-USB connector on it and batteries will typically have a USB-A ``output'' connector on them, thus a micro-USB to USB-A cable is needed.  Many times, this cable is included in the battery purchase.~\cite{cable}  Construction of the contact tracer consists of simply connecting the battery to the token.

\begin{figure}
\begin{center}
\includegraphics[scale=0.5]{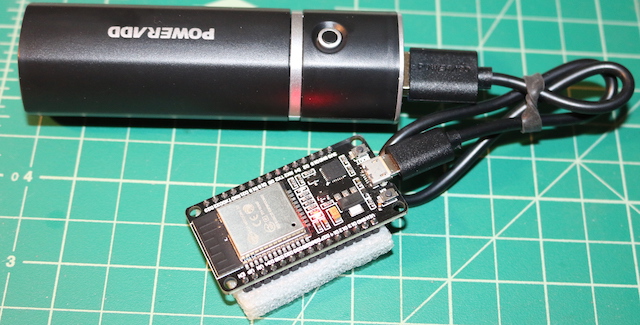}
\caption{Fully completed CT token described here.  The battery form may differ and the ESP32 (the circuitboard) should be enclosed in a protective and non-metallic case. See Ref.~\citenum{3dcase}.}
\label{ready_to_go}
\end{center}
\end{figure}

The token comes as a bare circuitboard that should be enclosed for protection.  Any non-metallic (handmade) case will do (plastic, cardboard, leather, canvas, cloth, etc.).  Custom-fit 3D-printable designs are available.~\cite{3dcase} The fully assembled CT (minus a case) is shown in Fig.~\ref{ready_to_go}.

\subsection{Flash token with contact tracer software}

The token needs to have firmware ``flashed'' onto it, so it may function as a contact tracer. This only needs to be done once and works as follows.

\begin{figure}
\begin{center}
\includegraphics[scale=0.35]{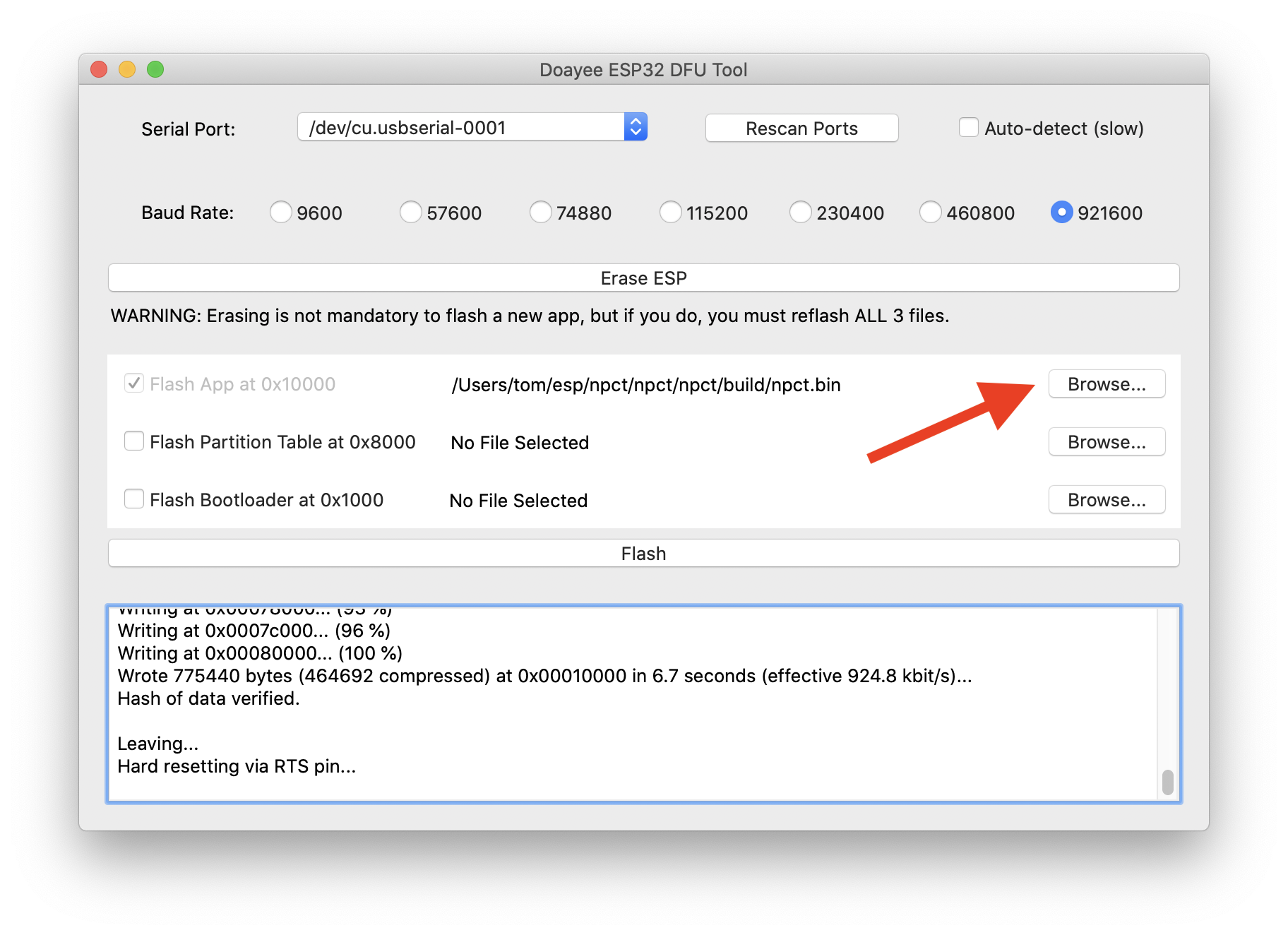}
\caption{Flashing software that ``flashes'' contact tracing software onto the token.}
\label{flash}
\end{center}
\end{figure}

First, download the CT firmware, which is a file called {\tt npct.bin.}~\cite{npct.bin} This is a pre-compiled binary that runs internally on the token. Flashing is the process of putting this file onto the token. To do so, download flashing software, available for both Windows and macOS, which will perform the one-time flashing.  We recommend {\tt DoayeeESP32DFU.app.zip} for macOS or {\tt DoayeeESP32DFU.exe} for Windows.~\cite{flash} When run, the software will look like that shown in Fig.~\ref{flash}. 

\begin{figure}
\begin{center}
\includegraphics[scale=0.25]{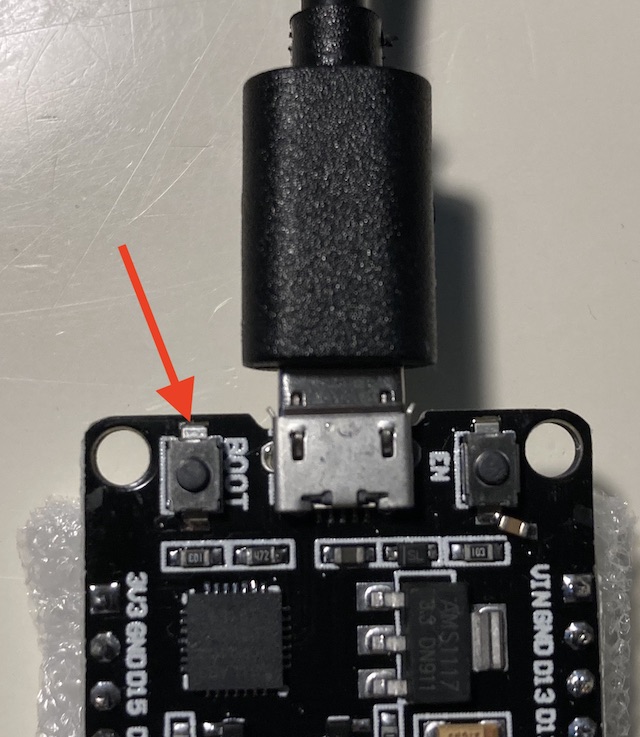}
\caption{Press this button to enable flashing of the contact tracing software.}
\label{boot_btn}
\end{center}
\end{figure}

Clicking the button indicated by the red arrow will allow one to navigate to the {\tt npct.bin} file previously downloaded.  Select this file, then click the large ``flash'' button.  Then, watch the large text box at the bottom for a message resembling {\tt "Connecting........\_\_\_\_\_... ."} When this is seen, press and hold the ``boot'' button on the ESP32, which is shown in Fig.~\ref{boot_btn}. 

If successful, writing percentages will begin to appear at which time the boot button may be released.  If nothing seems to happen or an error arises, try different serial port selections (top left dropdown in the flashing software of Fig.~\ref{flash}).

When this flash process completes, the {\tt npct.bin} file can be deleted, as can the Windows/macOS flashing software.

\subsection{Preparation for individual use in contact tracing}
\label{init}

At this point, the token is ready for everyday-use as a contact tracer. This means the token should go with the individual as they go out in public (placed in a pocket, bag, etc.)  First however, the device must be configured with the current health/symptoms profile of the individual.  This is done as follows.

\begin{figure}
\begin{center}
\includegraphics[scale=0.75]{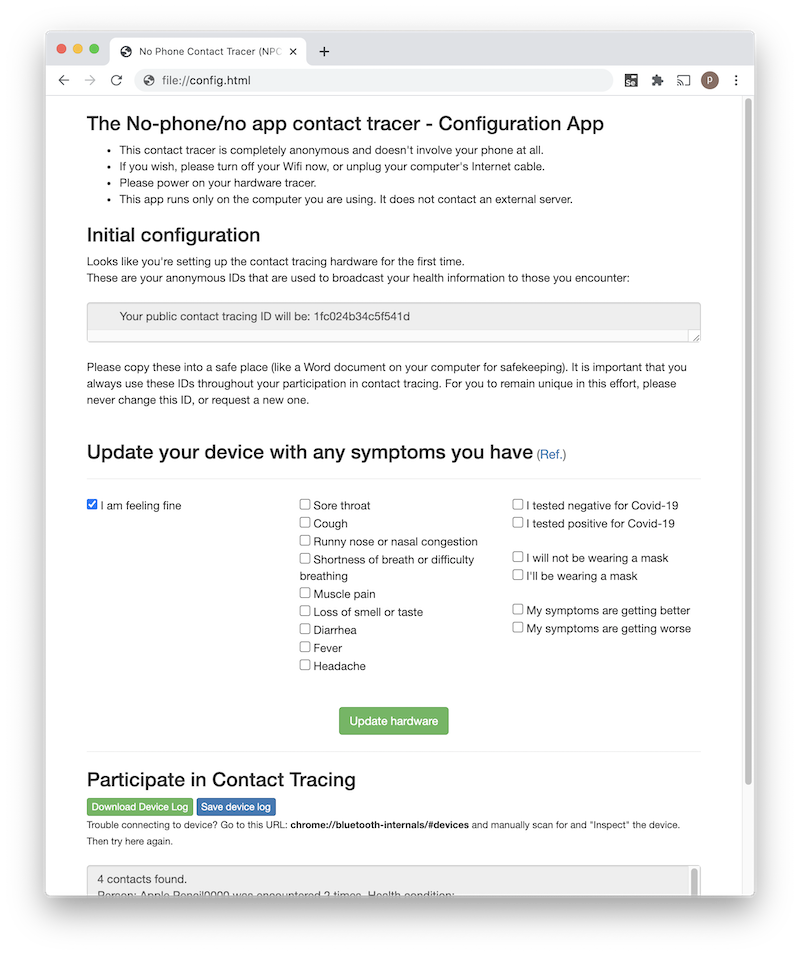}
\caption{The token's health configuration app.}
\label{configapp}
\end{center}
\end{figure}

In the project repository, there is a folder called  {\tt configapp},~\cite{configapp} containing a file called {\tt config.html}. Download this on the local computer and load it into a Chrome web-browser (version 83 or higher).~\cite{why_chrome} Do this by using the {\tt File$\rightarrow$Open File...} menu option. This will bring up a page in the browser that resembles that shown in Fig.~\ref{configapp}. We note that this Chrome-app is only an implementation convenience. The Internet is not  used at all. The suspect user may turn off the WiFi on their computer or disconnect any ethernet cable. This Chrome-app only communicates to the token via BLE.

The first time this Chrome-app is loaded, the user will be granted a one-time, anonymous public contact tracing identification (hereafter public-ID).~\cite{narayanan} This will be used throughout their participation in this CT program.  An example of such is shown here.
\begin{Verbatim}[frame=single]
Your public contact tracing ID will be: 2ef94e20ba20beea
\end{Verbatim}
No other account or personal information is required.  It is saved locally in storage associated with the Chrome browser, but are only accessible by Chrome itself.  A new public-ID is not generated in subsequent Chrome loads.  Thus, the same Chrome browser should not be shared amongst others configuring the token.

Next, the user ticks off self-prescribed symptoms, in this case related to COVID-19,~\cite{cdc_symptoms} then clicks the ``Update hardware'' button. This will update the token with the user's anonymous public-ID and health code, based on symptoms they are feeling.  A health code is formed by enumerating symptoms as per this list.

\begin{minipage}{\linewidth}
\begin{Verbatim}[frame=single]
1: "Feeling fine"
2: "Sore throat"
4: "Cough"
8: "Runny nose or nasal congestion"
16: "Shortness of breath or difficulty breathing"
32: "Muscle pain"
64: "Loss of smell or taste"
128: "Diarrhea"
256: "Fever"
512: "Headache"
1024: "Tested negative for Covid-19"
2048: "Tested positive for Covid-19"
4096: "Wearing a mask"
8192: "Not wearing a mask"
16384: "Symptoms are getting better"
32768: "Symptoms are getting worse"
 \end{Verbatim}
 \end{minipage}\\

A sum of all ticked symptoms is found and converted into a 4-digit hex code. So, for example a sore-throat and headache would sum to 514 or 0202 in hex. Thus, the complete datagram this user's token will share via BLE would be {\tt 2ef94e20ba20beea0202}.

This completes the configuration step of the token, which as mentioned, culminates with the user clicking the ``Update hardware''  button in the Chrome-app. These steps only need to be repeated as the user updates any symptoms. The need for the private verification code will be discussed later.

\section{Operational overview}

The user will power up the token by connecting it to the battery, and take it with them as they go out in public.  Software on the token is instructed to set the BLE name of the token using the prefix {\tt \#C19:} +  the public id + the health code.  Referring to the above example, the BLE name of this token will be {\tt \#C19:2ef94e20ba20beea0202}. When powered, software in the token will ``advertise'' this name using BLE, to all other tokens in proximity.  

For both the user (and those in proximity who are also carrying a similarly configured token), software (on the token) is also programmed to ``discover'' BLE names, and only log those having the {\tt \#C19:...} form above. Such a discovery is called an ``encounter.''  Encounters are held in the internal memory of a given user's token provided power is applied. 

When the user returns home, they may retrieve all logged encounters using the same Chrome-app as described above, using the ``Download device log'' button. This will show the public-ID and health codes of all encounters. In the interest of privacy, there is no automatic (online) saving of the downloaded encounters, but a button is offered to allow the user to save the encounters into a local file on their computer.

With the operational theme of this CT token discussed, we now move on to how it may help slow the current pandemic.

\section{Discussion on use in contact tracing}

\subsection{Conflicting COVID-19 issues}

We see two conflicting issues perpetuating the current pandemic as they may apply to CT. Issue one is is that as a guide, 50\% of people show symptoms of COVID-19 within 5 days of becoming infected (the ``50/5 time'').~\cite{jhbph_class}  This means for contact tracing to work, some intervention would have to occur within this time period.  

Working with this parameter of COVID-19 however is self-defeating issue two: lengthy times to both obtain a test and its result. In the author's own experience (certainly a function of locality), a test may be obtained only if one is showing symptoms, can take approximately a week for an appointment to have the test administered, then approximately 10 days to obtain the testing result.

\subsection{How this CT system may help}

It is thought this CT system can help in the following way. Although the clinical presentation of COVID-19 has a large variation,~\cite{jhbph_class_2} typical watch for symptoms appear to be somewhat regular.~\cite{watch_for_symptoms,jhbph_class} Suppose someone decides to be vigilant about their symptoms and is dedicated to using this system.  We see two benefits of its use.

First, it allows them one to discretely ``tell'' others they encounter about their symptoms. One's condition may start with a mild form of just one of the common symptoms (a slight cough for example). But, since the virus is still spreading as of August 2020, such a condition is obviously not enough cause individuals en mass to change their excursion plans and remain at home.  So, at least now others can know about their condition.

Second, another COVID-19 guide is that an elevated risk of infection occurs when there is physical, close, or proximate contact with one who is contagious.~\cite{jhbph_class_2}  Suppose the downloaded log reveals an encounter with someone in (at least) a similar health situation of showing {\sl some} symptom.  Obtaining this information would serve as a definitive point of reflection for the user considering the three elevated risk situations above.  

The user may now consider  what they were doing when the encounter came in. Which of the three encounter types likely occurred? Were they (and/or) any of the contacts were wearing a mask?  The token's log will also tell them the number of times each encounter happened, which is a measure of the contact intensity.  They may conclude that their own 50/5 time may now be underway.

Given these extraordinary times (and likely everyone's desire to emerge from this pandemic), the user might now consider changing their plans and lightly isolating for a couple of days, while they see if more symptoms appear or worsen. They could consider wearing a mask, even at home, and distance within their home (if possible). They would also keep careful track of any contacts, while having a call to their doctor or a testing station imminent.  Normal activities may resume only when something definitive comes along clearing them to do so.

\subsection{Self-motivated contact tracing}

This CT system admittedly relies heavily of the self-motivation and dedication from the user.  They must build, maintain, carry, and observe the resulting encounter log of the token.  We acknowledge that such widespread dedication is unlikely to occur.  However, the current pandemic is showing few signs of passing and we wonder if this system might draw from each individual to help us all emerge from it.~\cite{acdc} Experts have noted that our ``behavior must change,''~\cite{birx} and perhaps diligence with such a token could be a part of this. 

We note wearable form factors of other CT tokens.~\cite{hardware_token} The visibility of such may be a curious point for others to observe, namely they see someone who is actively participating in contact tracing. This may pique their own interest, in a similar manner to the ``I voted'' stickers worn during elections in the United States.~\cite{i_voted}

\section{Ongoing work: encounter sharing}

It is difficult to extend the functionality to this CT system given the insistence on privacy and security in its design (free of usual connectivity like WiFi, etc.). All functionality to this point is all contained between the user and the token. However, if (admittedly ``one more'' or ``just another'') a central web-based database it brought into the plan, additional benefits can be gained. This work is currently underway, and is summarized here.

\subsection{Anonymity, security and privacy}

A key security and privacy design aspect of this system is the sole public-ID mentioned in Section~\ref{init}.  It is a sequence of characters, completely unlinked to the individual and its uniqueness is all that is needed to participate in this peer-to-peer CT system.   Internally, the public-ID is actually derived from the private verification code forming a public/private key, similar to that used in message encryption.~\cite{narayanan}.  The private verification code can be used verify that the public-ID indeed originated from the initial Chrome-app session.

So, if we are satisfied with the core level of anonymity and the key-based security of this system, perhaps the {\sl sharing of encounters} could also be a part of this CT system. This is discussed below.

\subsection{Encounter sharing}

When a user downloads their encounter log from the token, they may choose to ``share'' their encounters with a central database. A given entry would resemble a line like
\begin{Verbatim}[frame=single]
2ef94e20ba20beea,8a04e24bcd91beea,2020-08-01
\end{Verbatim}
indicating that user {\tt 2ef94e20ba20beea} encountered {\tt 8a04e24bcd91beea} on 2020-August-01.  This would allow for a growing log of encounters. There are two prominent uses of such sharing. First, a user can post messages, which will be available to their encounters. Such messages could be about test results or degree of symptoms.  Second, suppose someone becomes symptomatic or even tests positive for COVID-19. Encounter sharing would provide a mechanism for this person to inform recent encounters of such.

Work is currently underway implementing these encounter sharing mechanisms.~\cite{npctshare} They will be integrated into the Chrome-app for ease of use. Optionality of this step will be emphasized and it will be clear how this sharing can be accomplished free of any log-ins, accounts, or divulging of any personal identifying information.

\section{Conclusions}

We have demonstrated an inexpensive, anonymous, peer-to-peer hardware contact tracer token that can be constructed by an individual from inexpensive parts.  It does not use mobile a phone or app at all, and is only optional in its use of a central database.  We acknowledge the burden placed on an individual to use this system, but wonder if the desire to emerge from this pandemic might help in this regard.  Using it would require a change in behavior involving diligent monitoring and sharing of both one's health and that of those they encounter.

Likely this contact tracing system might find its best use in (small) managed groups of people who share a common space, where a group leader could motivate its use.  This includes places of work, offices and schools. This system may help such entities to open safely and remain open.  In the case of schools, this system may serve as an ongoing and active student project on contact tracing. 

The author welcomes any discussions about this system.  Please contact at {\tt tbensky@calpoly.edu}.


\begin{thebibliography}{99}
\bibitem{frag} There are dozens of apps available, some written by individuals, companies, and even government entities.
\bibitem{why_chrome} Such versions of the Chrome browser must be used, since these support Web-Bluetooth.
\end{thebibliography}
\end{document}